# Reconfiguring Digital Accountability: AI-Powered Innovations and Transnational Governance in a Postnational Accounting Context

Claire Li
Royal Holloway, University of London

David Peter Wallis Freeborn
Northeastern University London

**Abstract**

This study explores how AI-powered digital innovations are reshaping organisational accountability in a transnational governance context. As AI systems increasingly mediate decision-making in domains such as auditing and financial reporting, traditional mechanisms of accountability, based on control, transparency, and auditability, are being destabilised. We integrate the Technology Acceptance Model (TAM), Actor-Network Theory (ANT), and institutional theory to examine how organisations adopt AI technologies in response to regulatory, ethical, and cultural pressures that transcend national boundaries. We argue that accountability is co-constructed within global socio-technical networks, shaped not only by user perceptions but also by governance logics and normative expectations. Extending TAM, we incorporate compliance and legitimacy as key factors in perceived usefulness and usability. Drawing on ANT, we reconceptualise accountability as a relational and emergent property of networked assemblages. We propose two organisational strategies including internal governance reconfiguration and external actor-network engagement to foster responsible, legitimate, and globally accepted AI adoption in the accounting domain.

**Keywords:** AI accountability, Transnational governance, Actor-Network Theory, Institutional pressures, Technology acceptance, Accounting regulation

# 1.0 Introduction

The accelerating integration of artificial intelligence (AI) into organisational processes raises critical questions about how accountability is constructed, distributed, and maintained in a transnational digital economy. While AI promises efficiency and innovation (Fountaine, McCarthy and Saleh, 2019; Olan et al., 2022), it simultaneously destabilises conventional mechanisms of organisational accountability, particularly within the domains of financial reporting, compliance, and governance. In this paper, we interrogate how AI-powered innovations reshape the landscape of digital accountability and explore the institutional, ethical, and cultural pressures that drive organisations to adopt such technologies. Our analysis is grounded in a transnational governance perspective (Djelic & Sahlin-Andersson, 2006; Roger & Dauvergne, 2016) and employs the Technology Acceptance Model (TAM) (Davis, Bagozzi and Warshaw, 1989; Silva, 2015) and Latourian Actor-Network Theory (ANT) (Latour, 1987, 2005) to examine how accountability is co-constructed across actors, institutions, and technologies.

AI adoption does not occur in a vacuum. Rather, it unfolds within socio-technical networks that are increasingly global, complex, and normatively contested (Jobin et al., 2019; Fazelpour and Danks, 2021). These networks comprise regulators, developers, situated practitioners, standard-setting bodies, and algorithmic systems, all interacting within diverse cultural and legal contexts (Toon, 2024). From financial disclosures shaped by AI-generated narratives to algorithmic auditing systems designed to perform compliance, accountability has become distributed across human and non-human actors (Matthias, 2004; Zednik, 2021). Existing frameworks, such as TAM, offer important insights into user-level adoption and perceived usefulness (Koenig-Lewis et al., 2015; Kim, Mirusmonov, & Lee, 2010), but they insufficiently capture how accountability and legitimacy are reconfigured in postnational accounting environments (Arnold, 2009a; Mehrpouya and Salles-Djelic, 2019).

Our study addresses two interrelated questions: (1) Why do organisations develop and adopt AI-powered digital innovations in the face of regulatory, ethical, and social pressures? and (2) How can organisations restructure their internal governance and external relationships to enhance accountability and legitimacy in transnational settings?

We propose a dual-theoretical approach to address these questions. First, we extend TAM by incorporating regulatory compliance, ethical principles, and cultural alignment as critical dimensions of perceived usefulness and ease of use (Kelly, Kaye, & Oviedo-Trespalacios, 2023; Pan et al., 2018). Second, we draw on ANT to conceptualise accountability not as a fixed organisational attribute but as a relational and emergent property of socio-technical networks (Latour, 2005; Djelic & Quack, 2007). ANT allows us to examine how AI accountability is enacted across boundaries, negotiated through governance assemblages, and contested within transnational regulatory regimes (Lobschat et al., 2021; Blanchard, Thomas, & Taddeo, 2024).

In addition to TAM and ANT, we also draw on insights from institutional theory to contextualise how social pressures shape AI adoption in transnational environments. Particularly, new institutionalist perspectives help illuminate how organisational behaviours are guided by coercive (legal), normative (ethical), and mimetic (peer-driven) pressures (DiMaggio & Powell, 1983). These pressures are embedded in global governance frameworks and professional expectations that increasingly define what constitutes "responsible" AI use. By incorporating institutional theory, we extend our conceptual framework to account not only for how accountability is constructed (via ANT), but also for why certain forms of AI-related accountability become dominant, legitimate, and institutionalised across organisational fields.

Social pressures including compliance with AI regulations (OECD, 2021; European Union, 2024), alignment with ethical frameworks (Ashok et al., 2022; Jobin et al., 2019), and responsiveness to stakeholder expectations (Lee, Pak and Roh, 2024), act as institutional motivators for technological transformation (Gegenhuber et al., 2022; Saarikko, Westergren and Blomquist, 2020). These pressures are particularly pronounced in accounting-related domains, where transparency, trust, and integrity are paramount (Djelic & Sahlin-Andersson, 2006; Arnold, 2009b). When organisations adopt AI-driven systems for decision-making, reporting, or compliance, they do so under public and regulatory scrutiny that transcends national jurisdictions.

Based on this framing, we propose two organisational responses: (1) internal reconfiguration of governance structures and resource capabilities to meet evolving standards of digital accountability (Song, Lee, and Khanna, 2016; Deloitte, 2024), and (2) strategic management of actor-networks that shape AI's legitimacy and market acceptance across borders (Latour, 2004; Pan et al., 2018). These responses are not merely operational strategies but normative interventions that challenge existing accountability arrangements and signal new modes of transnational responsibility.

Our study makes three key contributions to the literature on digital transformation, governance, and accountability.

First, we investigate why and how organisations develop and adopt AI-powered digital innovations from a transnational governance perspective. While prior research has examined the development of digital infrastructures and innovation processes (Tan, Abdaless and Liu, 2018; Tan, Liu and White, 2013), and explored market acceptance through organisational semiotics (Pan et al., 2018; Hafezieh and Eshraghian, 2022; Hafezieh and Pollock, 2023), we extend this stream by considering how organisations must adapt their innovations to cross-border legal, ethical, and cultural environments. We show that such contextual complexities reshape not only adoption decisions but also the ways in which accountability is structured and distributed.

Second, we foreground the role of institutional and normative pressures, particularly regulatory mandates, ethical expectations, and cultural legitimacy, in driving digital transformation. Previous literature has focused largely on functionalist factors such as ease of use, perceived usefulness, and security risks (Kim et al., 2010; Lu et al., 2011; Shin, 2010). Our study adds another layer by framing these pressures as forms of transnational social accountability that compel organisations to align their digital strategies with shared ethical standards and legal frameworks.

Third, we respond to Vial's (2021) call to examine how organisations improve market acceptance of digital transformation by introducing the concept of transnational accountability as a relational and dynamic construct. Using Actor-Network Theory (Latour, 2005), we theorise accountability as co-constructed across socio-technical systems and propose two organisational strategies to navigate this complexity: (1) internal reconfiguration of governance mechanisms, and (2) external management of actor-network relationships. These strategies challenge static understandings of accountability and offer a more nuanced, practice-oriented framework for responsible AI innovation in a postnational context.

## 2.0 Motivations of Developing and Applying AI-powered Digital Innovations

Organisations adopt AI-powered digital innovations for a variety of motivations, ranging from efficiency gains to strategic legitimacy in global markets. However, these motivations are not purely instrumental or technical; they are shaped by institutional, regulatory, and socio-cultural forces that increasingly transcend national boundaries. In this section, we explore three interrelated drivers of AI adoption: (1) economic incentives, (2) user acceptance logics grounded in the Technology Acceptance Model (TAM), and (3) institutional pressures emerging from transnational governance.

### 2.1 Profit-driven Motivations and the Reframing of Efficiency

Economic incentives remain a foundational motivation for organisations to adopt AI technologies. AI-powered systems enable automation of routine tasks, predictive analytics, and optimisation of decision-making processes, all of which contribute to enhanced profitability and operational efficiency (Fountaine, McCarthy and Saleh, 2019; Olan et al., 2022). These technologies offer not only cost reduction but also the potential for revenue growth through customer personalisation and service innovation (Usman et al., 2024).

However, when examined through an accountability lens, these economic motives also raise normative questions: who benefits from AI-enabled optimisation, and at what cost? In contexts such as financial reporting, internal control, or audit decision-making, efficiency gains may come at the expense of explainability, transparency, or ethical clarity. Thus, the pursuit of AI for profit is not neutral; it reconfigures organisational priorities and redistributes accountability between human actors and algorithmic agents.

### 2.2 Market Acceptance and the Technology Acceptance Model (TAM)

Understanding how organisations evaluate AI innovations requires attention to both instrumental utility and socio-legitimacy. The Technology Acceptance Model (TAM) (Davis, 1989; Silva, 2015) provides a foundational framework for explaining technology adoption, especially through the constructs of "Perceived Usefulness" and "Perceived Ease of Use". AI technologies are more likely to be accepted when they are seen to enhance job performance and when their interface and integration present minimal friction.

While TAM remains influential, its original formulation assumes that acceptance is largely shaped by individual cognition. In organisational and transnational contexts, however, acceptance is mediated by wider accountability structures. For example, in accounting functions, perceived usefulness may also hinge on whether an AI system produces auditable outcomes, complies with regulatory expectations, or protects client data integrity. Similarly, ease of use may depend on whether a system can be aligned with existing governance protocols and internal reporting standards.

By embedding TAM within broader accountability frameworks, we can better explain why certain AI innovations gain traction while others are resisted, not simply due to technical limitations, but because of perceived misalignments with ethical norms, control expectations, or professional responsibilities.

### 2.3 Extending TAM: Transnational Governance and Institutional Pressures

The global nature of AI deployment introduces institutional complexities that cannot be captured solely by user-level acceptance models. Recent research in accounting has highlighted the growing importance of transnational governance in shaping corporate behaviour and digital infrastructure (Arnold, 2009a; Mehrpouya and Salles-Djelic, 2019; Friedrich, Kunkel and Thiemann, 2024). Transnational governance refers to the diffuse yet coordinated efforts of states, international bodies, and private actors to establish rules, norms, and standards across jurisdictions (Djelic & Sahlin-Andersson, 2006; Roger & Dauvergne, 2016).

In the realm of AI, this includes formal regulations such as the EU AI Act (European Union, 2024), soft-law instruments like the OECD AI Principles (OECD, 2021), and emerging ethical frameworks promoted by industry alliances and civil society. These governance regimes do not merely prescribe compliance; they reshape how accountability is understood and enacted. For instance, they compel organisations to demonstrate responsible data usage, transparency in algorithmic decision-making, and fairness in outcomes, often across vastly different legal and cultural settings.

We argue that TAM can be extended to include these institutional variables. In a transnational setting, "Perceived Usefulness" encompasses not only functional outcomes but also compliance capability, i.e., the extent to which a technology supports alignment with global legal and ethical expectations. "Perceived Ease of Use" similarly includes the adaptability of a system to different cultural, regulatory, or reporting environments.

By integrating governance considerations into technology acceptance frameworks, we foreground the fact that AI adoption is not only about user preferences or market logic, but about navigating complex assemblages of accountability, legitimacy, and institutional constraint.

### 2.4 Social Pressures as Institutional Motivations

Beyond economic or technical logics, the adoption of AI-powered digital innovations is increasingly shaped by institutional pressures embedded in transnational governance environments.

Drawing on new institutional theory (DiMaggio & Powell, 1983), we frame these pressures as comprising three interrelated forms: coercive, normative, and mimetic. These forces create not only incentives but also constraints that guide organisations toward field-level conformity around responsible AI governance, particularly in accounting-related domains where transparency and trust are paramount (Arnold, 2009a; Djelic & Sahlin-Andersson, 2006; Mehrpouya & Salles-Djelic, 2019).

Table 1 summarises key transnational AI governance instruments and their implications for accounting-related practices. These instruments vary in legal enforceability and design philosophy, ranging from strict, rules-based regimes like the EU AI Act to principles-based guidelines such as those issued by the OECD and UK Government. Each framework foregrounds different aspects of AI governance - risk, fairness, explainability, or professional ethics - which, in turn, shape how accounting organisations configure their systems for compliance, legitimacy, and accountability.

**Table 1: Selected Transnational AI Governance Instruments and Their Relevance to Accounting Accountability**

| Regulation/Policy | Region/Organisation | Type | Core Elements | Relevance to Accounting Practice |
|---|---|---|---|---|
| EU AI Act (2024) | European Union | Rules-based | Risk categorisation, mandatory transparency, algorithm registry | Audit automation, internal controls, traceable data pipelines |
| OECD AI Principles | OECD Member States | Principles-based | Fairness, transparency, accountability, robustness | Supports cross-border alignment of reporting and assurance norms |
| UK AI White Paper (2023) | United Kingdom | Principles-based | Outcome-driven, flexible, focused on safety and fairness | Applicable to global audit frameworks with flexible AI components |
| GDPR | European Economic Area | Legal regulation | Data privacy, lawful use, portability | Financial and employee data compliance |
| IFAC AI Ethics Exposure Draft (2023) | IFAC/IAASB | Professional guidance | Professional ethics, auditability, responsibility chain | Accountability in AI-assisted auditing and reporting |
| World Bank AI Governance Toolkit (2024) | Global/Multilateral | Framework | Regulatory capacity, risk oversight, cross-jurisdictional norms | Relevant for transnational disclosures in developing economies |

From a TAM perspective, these frameworks influence *Perceived Usefulness* by defining what AI systems are deemed "effective" or "acceptable" under legal and ethical scrutiny. For example, an AI system that satisfies the EU AI Act's transparency standards may be seen as more trustworthy and functionally beneficial in audit settings. Similarly, *Perceived Ease of Use* is affected by whether an AI tool is interoperable across legal jurisdictions, i.e., whether it can easily adapt to both GDPR and OECD expectations.

From an ANT lens, these regulatory instruments are not external constraints but actors within the network of accountability; they co-shape how governance is enacted, how roles are distributed between developers, accountants, and regulators, and how legitimacy is performed across regions.

**2.4.1 Coercive Pressures: Regulation and Risk Governance**

Coercive pressures arise from formal legal and regulatory requirements imposed by states, regional blocs, and transnational institutions. The proliferation of AI regulations across jurisdictions such as the European Commission AI Act (European Union, 2024), the OECD AI Principles (OECD, 2021), and emerging frameworks in the UK (UK Government, 2023), has created a dense regulatory environment that directly conditions AI innovation and implementation. These frameworks emphasise issues such as data privacy, algorithmic transparency, and accountability in AI development and use (Fazelpour & Danks, 2021; Floridi & Cowls, 2019).

Regulatory regimes differ in logic and application. Rules-based systems, like the EU AI Act, offer legal clarity and enforcement but may be rigid and slow to adapt to technological change. Principles-based frameworks, like the UK White Paper, enable flexibility and innovation but pose challenges in interpretation and compliance consistency (Ashok et al., 2022; Blanchard et al., 2024). As a result, organisations operating globally must navigate cross-jurisdictional uncertainty to avoid reputational damage, legal sanctions, and loss of consumer trust (Bursztyn & Jensen, 2017; World Bank Group, 2024).

In this sense, regulation becomes a "coercive social pressure" and a key site where compliance, legitimacy, and digital accountability intersect (Lobschat et al., 2021; Song, Lee, & Khanna, 2016).

Regulatory compliance is especially complex in accounting contexts, where AI-powered systems are used in risk-based auditing, financial forecasting, fraud detection, and ESG performance reporting. The application of AI in these domains must not only comply with AI-specific laws (e.g., GDPR, EU AI Act) but also with sector-specific regulations such as IFRS disclosure requirements, SOX internal control rules, or GRI sustainability standards. These regimes often demand traceability, auditability, and human accountability, criteria that many AI systems currently struggle to meet (Zednik, 2021; Deloitte, 2024).

However, transnational regulatory frameworks are not ideologically neutral. Governance instruments such as the EU AI Act function as vehicles for exporting specific normative assumptions, often grounded in Western legal traditions, into global digital infrastructures. This raises concerns over regulatory asymmetry, where jurisdictions in the Global South or non-Western contexts must conform to standards they did not co-produce. Such dynamics reflect an emergent politics of AI governance, where power is exercised through standard-setting and compliance conditionality, rather than formal legislation alone.

This layering of regulatory obligations produces what Power (2015) calls "infrastructural accretion," where accountability becomes embedded in overlapping technical, legal, and institutional systems. Organisations that fail to align their AI innovations with this multilayered regulatory environment risk not only non-compliance but also erosion of legitimacy within global financial ecosystems.

In addition to regulatory asymmetry, transnational AI governance reflects ongoing struggles over norm-setting authority. For instance, while the EU advances strict rules-based AI frameworks, other regimes such as China or the US adopt more flexible, innovation-first approaches. These tensions not only affect interoperability but also risk fragmenting the global AI accountability landscape, raising the possibility of competing governance blocs with divergent enforcement logics.

**2.4.2 Normative Pressures: Ethical and Cultural Alignment**

Normative pressures stem from societal values, ethical norms, and stakeholder expectations around responsible AI. Societies increasingly demand that AI systems be designed and deployed in ways that ensure fairness, non-discrimination, safety, transparency, and respect for privacy (Jobin et al., 2019; Johnson, 2020; Zednik, 2021). These expectations are formalised in industry standards and codes but are also culturally situated and often contested (Sanderson et al., 2023).

The core ethical principles frequently cited in global AI governance - transparency, fairness and equity, non-maleficence, responsibility, and privacy - represent aspirational norms, but their implementation often involves tensions and trade-offs (Nissenbaum, 2010; Matthias, 2004). For example, fairness in algorithmic decision-making may reduce accuracy, and transparency may conflict with privacy, especially in accounting algorithms used for credit scoring, tax compliance, or risk profiling (Fazelpour & Danks, 2021; Zednik, 2021).

Such normative tensions are not merely theoretical. For instance, AI systems used in ESG performance analysis may prioritise fairness and environmental accountability, using demographic or community-level data to assess social impact. However, the same variables may be legally restricted or discouraged in algorithmic credit scoring models due to privacy or anti-discrimination laws in financial regulation. This conflict creates a governance paradox: an AI tool that enhances accountability in sustainability reporting may simultaneously violate principles of fairness or legality in consumer finance. These tensions illustrate how normative expectations can diverge across application domains, demanding more context-sensitive approaches to AI governance in accounting and financial contexts.

Companies that align their AI innovations with these principles are more likely to be perceived as socially responsible and trustworthy, which enhances public legitimacy and stakeholder acceptance (Lobschat et al., 2021; Ko & Leem, 2021). Ethical compliance thus operates as a "normative social pressure", particularly salient in fields like financial disclosure, automated auditing, and ESG reporting where public trust is critical (Athota et al., 2023; Deloitte, 2024).

These expectations are increasingly codified into ethical frameworks across jurisdictions. Jobin et al. (2019), in their comparative study of 84 AI ethics guidelines, identify five commonly recurring principles: transparency, fairness, non-maleficence, responsibility, and privacy. While tensions exist between these principles in practice (Blanchard et al., 2024; Sanderson et al., 2023), their repeated appearance suggests an emerging transnational normative consensus. For organisations operating across borders, aligning with these principles becomes not only an ethical imperative but also a strategy for securing stakeholder legitimacy and accountability.

Beyond abstract ethical principles, the cultural embeddedness of AI systems poses a distinct set of challenges. What is considered "fair" or "accountable" in one jurisdiction may be interpreted differently in another due to divergent historical, religious, and political traditions (Toon, 2024; Fazelpour & Danks, 2021). For instance, Western models of algorithmic transparency may prioritise individual autonomy, while East Asian contexts may emphasise social harmony or institutional trust. These differences shape how organisations localise AI applications and interpret ethical expectations.

In transnational accounting practices, such cultural divergences become particularly salient. An AI tool designed to assess ESG disclosures in a European context may rely on assumptions about materiality or stakeholder relevance that do not hold in other regions. This raises the risk of normative dissonance, where AI-driven accountability systems reflect culturally partial logics while claiming universality. As such, cultural adaptation is not merely a localisation task; it is central to the legitimacy of transnational digital accountability regimes.

### 2.4.3 Mimetic Pressures: Field-level Imitation and Competitive Signalling

Mimetic pressures emerge when organisations adopt innovations in response to perceived best practices or institutional uncertainty. As AI becomes institutionalised in corporate governance and reporting, organisations often imitate early adopters or industry leaders to maintain legitimacy and avoid being seen as laggards (Gegenhuber et al., 2022; Pan et al., 2018). This is especially relevant in highly visible domains like audit automation, narrative reporting, or ethics-driven AI innovation platforms (Hafezieh & Eshraghian, 2022; Hafezieh & Pollock, 2023).

Participation in global governance forums such as the UN's AI for Good Summit or the Partnership on AI, further allows firms to signal commitment to responsible innovation and shape emerging standards (OECD, 2021; Roger & Dauvergne, 2016). Such imitation reflects not technical necessity but a desire for symbolic alignment with perceived legitimacy norms, especially in transnational industries such as accounting, finance, and digital compliance.

Together, these three forms of institutional pressure, i.e., coercive, normative, and mimetic, shape how organisations adopt AI-powered innovations. They motivate the restructuring of governance practices, the embedding of ethical standards, and the pursuit of field-level legitimacy. From a transnational governance perspective, these pressures do not only accelerate technology adoption; they also create conditions for "distributed and mutual accountability" across borders, professions, and systems (Djelic & Quack, 2007; Djelic & Sahlin-Andersson, 2006).

In this light, social pressures are not peripheral, they are "constitutive drivers" of digital innovation in postnational contexts, particularly in accounting where auditability, explainability, and legitimacy are non-negotiable.

Transnational accountability, in this context, refers to the distributed responsibility that organisations bear toward multiple, overlapping regulatory regimes and stakeholder communities across borders. Unlike traditional models of accountability that are vertically assigned within nation-states, transnational accountability emerges relationally through interactions between organisations, global governance institutions, industry standards bodies, and civil society actors (Djelic & Quack, 2007). It is not static but constantly negotiated, shaped by legal compliance, ethical alignment, and reputational legitimacy in a pluralistic, postnational governance environment.

Figure 1 illustrates how institutional pressures operating within transnational governance environments shape organisational responses to AI accountability. The top layer represents the institutional environment - regulatory frameworks, ethical norms, and governance initiatives - that collectively define the emerging accountability expectations. These exert coercive (legal), normative (ethical/cultural), and mimetic (symbolic/competitive) pressures on organisations. In response, organisations restructure internal governance mechanisms and strategically manage external stakeholder networks, resulting in the co-construction of transnational accountability. This framework conceptualises accountability not as a fixed compliance outcome, but as a dynamic process embedded in institutionalised governance assemblages.

**Figure 1: Institutional Pressures Driving Transnational AI Accountability**

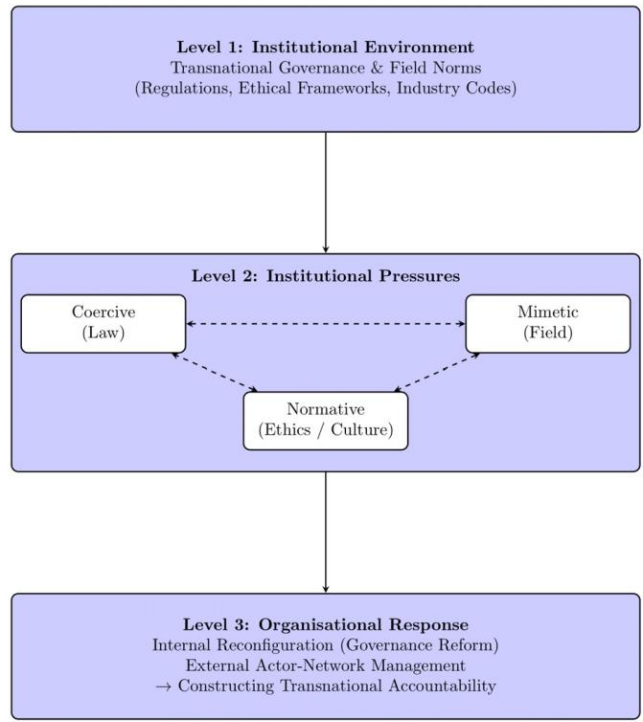

These institutional pressures are not always aligned. For example, coercive pressure for traceability (e.g., under SOX or GDPR) may conflict with mimetic pressure to adopt opaque yet efficient AI audit tools used by industry leaders. Similarly, normative pushes for privacy may undermine transparency mandates in credit assessment AI systems. Such internal contradictions demand that organisations engage in continuous ethical prioritisation and governance trade-off management.

## 3.0 Proposed solutions: Reconstructing Accountability in Transnational AI Governance

This section develops two interrelated organisational strategies to enhance digital accountability in AI-powered innovations: internal reconfiguration and external actor-network management. These solutions are grounded in a triadic theoretical framework that integrates the Technology Acceptance Model (TAM), Actor-Network Theory (ANT), and institutional perspectives on transnational governance.

### 3.1 Theoretical Foundations: Reframing Accountability through TAM, ANT, and Transnational Governance

To navigate the complexity of AI accountability, we propose an integrated framework that brings together three complementary theoretical lenses.

First, the Technology Acceptance Model (TAM) (Davis, 1989) explains technology adoption in terms of "Perceived Usefulness" and "Perceived Ease of Use". In a transnational context, these constructs must be extended to include institutional concerns: usefulness now entails regulatory compatibility and stakeholder trust, while ease of use involves interoperability with legal, ethical, and cultural norms across jurisdictions (Kelly, Kaye & Oviedo-Trespalacios, 2023; Pan et al., 2018).

Second, Actor-Network Theory (ANT) (Latour, 2005) shifts the focus from internal decision-making to the socio-technical assemblages in which AI is embedded. ANT treats accountability not as a top-down mechanism but as an emergent property of interactions among humans, machines, data infrastructures, and regulatory agents. From this view, organisations are not isolated agents of innovation but participants in heterogeneous networks of responsibility, where governance is continuously negotiated and reassembled (Latour, 2004; Djelic & Quack, 2007). While TAM approaches AI adoption from a functionalist and individual-level perspective, ANT reframes adoption as an emergent, networked process that reconfigures organisational relationships and accountability logics. This theoretical tension enriches our analysis by combining behavioural acceptance models with structural and relational critiques of governance.

Third, transnational governance theory conceptualises accountability within globalised institutional architectures. It foregrounds the role of supranational standards (e.g., EU AI Act, OECD Principles), legal pluralism, and soft law in shaping organisational legitimacy across borders (Djelic & Sahlin-Andersson, 2006). These regimes do not merely impose rules; they define the conditions under which AI practices are accepted, challenged, or legitimised (Arnold, 2009b; Mehrpouya & Salles-Djelic, 2019).

Together, these frameworks inform our dual-strategy approach: internal transformation of organisational structures and practices, and external management of relational networks, both aimed at reconstructing accountability under conditions of institutional complexity and transnational scrutiny.

While each of the three theoretical lenses - TAM, ANT, and institutional theory - offers unique insights, their underlying assumptions differ in meaningful ways. TAM explains technology adoption in rationalist, user-centred terms, focusing on intention and perception. In contrast, ANT displaces the human user as the central actor and instead highlights how technologies, policies, and people co-produce each other through ongoing networked negotiations. Institutional theory adds a macro-level dimension, emphasising how organisations adopt technologies to maintain legitimacy under coercive, normative, or mimetic pressures. By juxtaposing TAM's functionalist view of acceptance with ANT's relational construction of accountability, we reveal the limitations of behavioural models in capturing how AI adoption unfolds within complex institutional and governance ecosystems. These theoretical tensions are not contradictions but productive frictions that deepen our understanding of AI as both an organisational artefact and a socially embedded governance challenge.

**3.2 Internal Reconfiguration: Governance Alignment and Ethical Capability Building**

A central strategy for enhancing transnational accountability in AI-powered innovations lies in internal organisational reconfiguration. This involves the realignment of governance structures, ethical oversight mechanisms, and workforce capabilities to meet the increasingly complex and overlapping regulatory, normative, and stakeholder demands. From the perspective of Actor-Network Theory (ANT), internal structures are not passive carriers of innovation but active participants in the co-construction of accountability (Latour, 2005). Organisations must therefore treat AI adoption not as a technical decision but as a "matter of concern" (Latour, 2004), requiring ongoing ethical reflexivity and procedural adaptation. In this sense, accountability becomes a design choice, shaped by how data are handled, algorithms are governed, and oversight is institutionalised.

This internal transformation also aligns with the Technology Acceptance Model (TAM), which explains how perceptions of usefulness and ease of use affect the uptake of technologies. In the context of transnational AI systems, these perceptions extend beyond situated practitioners to include regulators, auditors, and global stakeholders. By embedding fairness, transparency, and traceability into AI development processes (Ko and Leem, 2021; Kelly, Kaye, & Oviedo-Trespalacios, 2023), organisations not only improve trust but also reduce friction in regulatory approval and stakeholder acceptance. Reconfiguring internal systems to align with global AI governance regimes such as the EU AI Act, GDPR, and OECD AI Principles allows organisations to move from reactive compliance toward proactive legitimacy construction (European Union, 2024; OECD, 2021).

Developing internal capabilities in AI ethics, compliance, and communication is central to this transformation. Organisations must cultivate a workforce capable of navigating diverse ethical standards, legal systems, and cultural expectations. This includes rethinking internal reporting mechanisms to ensure auditability and explainability, particularly in accounting and financial systems where algorithmic decisions can materially affect reporting outcomes or stakeholder interpretation. These developments expose tensions between emerging AI governance regimes and legacy accounting standards. While the EU's AI Act demands explainability and algorithmic transparency, the IFRS does not yet mandate disclosure of algorithmic processes used in ESG or risk assessments, creating normative dissonance in AI-assisted reporting environments. As AI continues to shape financial infrastructures, these conflicts will intensify unless standards evolve to accommodate algorithmic reasoning. As noted by Song, Lee, and Khanna (2016), internal co-evolution of technical and ethical systems enhances long-term resilience. Moreover, effective resource optimisation accelerates AI deployment while ensuring alignment with global compliance frameworks, improving both competitiveness and public trust (Deloitte, 2024).

In highly fragmented transnational environments, where legal and ethical standards vary across jurisdictions, the capacity for internal flexibility becomes a strategic advantage. Rather than seeking to standardise governance globally, organisations must construct adaptable internal architectures capable of shifting between accountability logics: legal compliance in one region, stakeholder engagement in another, and ethical assurance in a third. Through this form of governance reconfiguration, organisations construct a reflexive, distributed accountability regime that anticipates not only technological risks but the evolving expectations of regulators, consumers, and professional institutions alike.

### 3.3 Reshaping Boundaries and Managing Actor-Network Dynamics

A second strategic response to the demands of transnational accountability lies in reshaping organisational boundaries and managing the dynamic actor-networks in which AI innovations are embedded. From an Actor-Network Theory (ANT) perspective, technologies are not neutral tools but participants in distributed socio-technical assemblages (Latour, 2005). In this framework, accountability is not pre-assigned to discrete entities but is co-produced through ongoing interactions among diverse actors - regulatory bodies, civil society organisations, technical platforms, end-users, and legal infrastructures. These interactions occur across institutional, cultural, and jurisdictional divides, making organisational boundary work a critical governance activity (Djelic & Quack, 2007).

To operate effectively in such settings, organisations must move beyond siloed innovation strategies and engage directly in collaborative governance processes. This entails participating in transnational forums such as the UN's AI for Good Summit, the OECD's Global Partnership on AI, and industry-specific multi-stakeholder platforms. These arenas allow firms not only to align with evolving ethical standards but also to influence emerging norms, anticipate policy shifts, and build symbolic legitimacy (OECD, 2021; Mehrpouya & Salles-Djelic, 2019). Such participation is not merely reputational; it is strategic, enabling organisations to shape the very infrastructures through which accountability is assigned and contested.

Managing actor-network dynamics also involves cultivating alliances across epistemic communities. Legal experts, software engineers, auditors, data ethicists, and financial controllers all bring competing perspectives on what counts as "responsible" AI. Organisations must orchestrate these heterogeneous inputs into coherent governance responses. This may involve co-developing explainability protocols with auditors, co-designing fairness metrics with affected communities, or engaging with local regulators to adapt global AI tools to regional norms (Floridi & Cowls, 2019; Jobin et al., 2019).

Strategic partnerships likewise play a central role. By collaborating with NGOs, standard-setting bodies, and academic institutions, organisations expand their capacity to navigate global accountability regimes. For example, AI tools used in cross-border financial reporting must meet not only technical robustness criteria but also satisfy differing cultural expectations around disclosure, materiality, and risk tolerance (Zednik, 2021; Sanderson et al., 2023). Through boundary-spanning engagements, firms become participants in a global governance ecosystem rather than passive rule-takers.

This proactive mode of actor-network management allows organisations to anticipate regulatory developments, reduce implementation friction, and co-construct shared ethical frameworks. It contrasts with reactive or minimalist compliance models, which often overlook the relational, negotiated, and situated nature of accountability. In volatile governance environments, legitimacy is not granted; it is continuously performed through participation, reflexivity, and alignment with distributed expectations.

For accounting-related AI applications, governance is not limited to legal regulators but also includes standard-setting and professional bodies such as the International Federation of Accountants (IFAC), the International Auditing and Assurance Standards Board (IAASB), and the OECD's Working Party on Responsible Business Conduct. These actors help shape the normative expectations around AI's role in financial reporting, audit assurance, and ethical disclosure. Together, they form a distributed governance assemblage that influences what constitutes legitimate, responsible, and auditable AI practices in transnational accounting fields.

In sum, managing actor-network dynamics is not about extending control over external entities but about reconfiguring how organisational responsibility is framed and enacted in a transnational field. By combining internal reconfiguration with external engagement, organisations can position themselves as *responsible AI stewards* - agents who not only adopt emerging technologies, but also help shape the institutional architectures within which those technologies are made accountable.

Transnational accountability is not achieved solely through regulatory compliance, but through the cultivation of perceived legitimacy within diverse stakeholder communities. This form of legitimacy often hinges on broader societal expectations of fairness, transparency, and inclusivity, expectations that cannot be fully captured by formal regulation alone.

**3.4 Implications for Accounting Accountability**

The rise of AI-powered innovations poses significant implications for how accountability is conceptualised and enacted within accounting domains. Traditional accountability in accounting has been structured around clear chains of responsibility, auditable records, and standardised disclosure frameworks. However, AI systems, particularly those involving machine learning, natural language processing, or autonomous decision-making, challenge these assumptions by introducing opacity, distributed agency, and algorithmic mediation into financial and governance processes (Zednik, 2021; Arnold, 2009a).

AI-powered innovation is not simply reconfiguring technical workflows; it is redefining the institutional grammar of accountability in financial governance. The introduction of machine learning and algorithmic systems into accounting processes challenges existing assumptions about responsibility, auditability, and legitimacy, necessitating a fundamental rethinking of what constitutes credible financial information and oversight.

Our analysis suggests that transnational AI accountability redefines the spatial and relational boundaries of accounting responsibility. Rather than being confined within organisational hierarchies or national regulatory frameworks, accountability now emerges through interactions between AI developers, data engineers, compliance officers, auditors, and transnational standard-setters. This dynamic has two major consequences for accounting practice.

First, it calls for a shift from "retrospective accountability" based on record-keeping and reporting toward "prospective and procedural accountability", where responsibility is embedded in design choices, algorithmic oversight mechanisms, and stakeholder engagement processes (Matthias, 2004; Mehrpouya & Salles-Djelic, 2019). For example, AI systems used in risk-based audit analytics must incorporate explainability protocols and ethical review checkpoints to ensure traceability and legitimacy across jurisdictions. For example, an AI-based ESG analytics platform designed to align with European non-financial disclosure directives may not satisfy U.S. GAAP's materiality thresholds or enforcement logic. Similarly, IFRS does not currently mandate algorithmic traceability or model auditability, raising questions about the admissibility of AI-generated figures in formal disclosures. These mismatches illustrate the regulatory friction that arises when AI systems are deployed across jurisdictions with divergent accounting norms, creating uncertainty in both professional practice and assurance standards.

Second, it introduces a new layer of "infrastructural accountability", whereby accounting professionals must not only interpret AI-generated outputs but also critically assess the governance frameworks within which those outputs are produced. This includes evaluating the adequacy of data inputs, the fairness of algorithmic classifications, and the compatibility of AI systems with local reporting standards or global ESG frameworks (Power, 2015; Floridi & Cowls, 2019). In this context, the accountant is no longer a neutral user of information systems, but an actor who co-produces the conditions of AI legitimacy and operational accountability.

Transnational governance intensifies these challenges by exposing accounting practices to pluralistic norms and divergent stakeholder expectations. What is deemed a legitimate AI-enhanced reporting tool in one country may be considered biased or non-compliant in another. Navigating such complexity demands that accounting professionals engage in reflexive boundary work, not only interpreting AI results but also shaping the very systems through which accountability is claimed and contested.

Overall, AI innovations compel the accounting field to rethink its foundational logics of responsibility, assurance, and governance. Rather than viewing AI as a threat to professional judgement, our framework positions it as a catalyst for evolving accountability infrastructures where human expertise, machine capabilities, and institutional logics are dynamically interwoven. For accounting scholars and practitioners, this shift presents both a challenge and an opportunity: to develop new conceptual models, ethical standards, and professional competencies that can sustain meaningful accountability in an increasingly algorithmic world.

### 3.5 Accounting Standard-Setting as Transnational AI Governance

Beyond formal regulation, AI adoption in accounting is increasingly governed by the implicit authority of standard-setting bodies such as the International Federation of Accountants (IFAC), the International Auditing and Assurance Standards Board (IAASB), and global institutions like the OECD. These actors function as transnational governors by shaping professional norms, audit expectations, and definitions of responsibility through their guidance on digitalisation, ethics, and assurance practices.

As AI tools begin to mediate reporting, auditing, and disclosure tasks, their design and use are subtly regulated through these standards—even in the absence of formal AI-specific mandates. This shift marks a form of indirect governance, where AI system capabilities are filtered through evolving interpretations of what constitutes "reasonable assurance" or "materiality". Consequently, the accounting profession finds itself not only responding to technological change but also co-producing the normative boundaries within which AI legitimacy is conferred.

This process reverses traditional assumptions about the passive role of technology in professional practice. Instead, AI becomes a force that reshapes professional judgement, decision pathways, and institutional expectations, redefining the contours of accountability within global financial governance.

## 4.0 Conclusion

This study has examined the organisational development and adoption of AI-powered digital innovations through an integrated theoretical lens, combining the Technology Acceptance Model (TAM), Actor-Network Theory (ANT), and transnational governance frameworks. Moving beyond conventional models of technology uptake, we have framed AI adoption as a response to layered institutional pressures including regulatory mandates, ethical norms, and cultural expectations that coalesce within global socio-technical networks.

Rather than viewing AI systems as inherently biased or autonomous actors, we position their accountability as contingent on how organisations design, deploy, and govern them. We argue that in a transnational context, accountability is not merely assigned; it is actively co-constructed across distributed networks of actors, including developers, regulators, auditors, civil society groups, and standard-setting bodies. This calls for a rethinking of traditional governance logics: accountability is no longer confined within national legal regimes or organisational hierarchies but emerges through interaction, negotiation, and alignment across institutional and jurisdictional boundaries.

Our contribution is threefold. First, we reconceptualise AI adoption not as a function of individual perception or legitimation effort alone, but as a response to institutionalised pressures that shape what technologies are considered acceptable, responsible, and legitimate. Second, we offer two interconnected organisational strategies - internal reconfiguration and actor-network management - that allow firms to navigate this complex environment. Third, we highlight the specific implications for accounting accountability, where AI systems are increasingly integrated into financial reporting, audit, and compliance functions, requiring new forms of ethical oversight, infrastructural adaptation, and professional judgement.

Practically, our framework encourages organisations to move beyond compliance-as-minimum and instead engage in reflexive accountability-building. This includes embedding ethical values into AI design, aligning internal governance structures with evolving transnational standards, and participating in multi-stakeholder governance arenas that shape the field of responsible AI innovation.

Future research can extend our conceptual framework in several directions. First, empirical studies could test the operationalisation of transnational accountability across different sectors and jurisdictions. Second, further investigation is needed into how internal organisational infrastructures accrete and evolve in response to competing regulatory and ethical demands (Power, 2015). Third, scholars could examine the mechanisms by which organisations manage actor-network tensions, including resistance, misalignment, and co-optation. Fourth, the relationship between market acceptance and consumer affordance (El Amri & Akrout, 2020; Hafezieh & Eshraghian, 2017) warrants deeper exploration, particularly in AI systems deployed for public-sector accountability or inclusive finance. Fifth, while our study identifies strategic responses conceptually, future work should translate these into actionable frameworks, policies, and toolkits through which organisations can realise ethical, legal, and cultural accountability in practice.

Finally, we acknowledge that not all AI technologies are the same. Symbolic or rules-based AI systems allow for predefined control but offer limited adaptability. Neural or generative models, by contrast, function as opaque and continuously learning systems - posing greater risks for accountability, auditability, and legal liability (Zednik, 2021). Understanding how different AI architectures interact with governance mechanisms remains an urgent area of inquiry. As AI continues to shape the infrastructure of accounting and organisational decision-making, scholars and practitioners alike must engage critically with how accountability is being redefined, not only in theory but also in practice, across borders, institutions, and systems.

# References

Arnold, P. J. (2009a). Global financial crisis: The challenge to accounting research. *Accounting, Organizations and Society*, 34(6-7), 803-809.

Arnold, P. J. (2009b). Institutional perspectives on the internationalization of accounting. In C. S. Chapman, D. J. Cooper, & P. B. Miller (Eds.), Accounting, Organizations & Institutions: Essays in Honour of Anthony Hopwood (pp. 47-64). Oxford University Press.

Arvidsson, N. (2014). Consumer attitudes on mobile payment services–results from a proof of concept test. *International Journal of Bank Marketing*, 32(2), 150-170.

Ashok, M., Madan, R., Joha, A., & Sivarajah, U. (2022). Ethical framework for Artificial Intelligence and Digital technologies. *International Journal of Information Management*, *62*, 102433.

Athota, V. S., Pereira, V., Hasan, Z., Vaz, D., Laker, B., & Reppas, D. (2023). Overcoming financial planners' cognitive biases through digitalization: A qualitative study. *Journal of Business Research*, *154*, 113291.

Bharadwaj, A., El Sawy, O., Pavlou, P., & Venkatraman, N. (2013). Digital business strategy: toward a next generation of insights. *MIS Quarterly*, 37(2), 471-482.

Blanchard, A., Thomas, C., & Taddeo, M. (2024). Ethical governance of artificial intelligence for defence: Normative tradeoffs for principle to practice guidance. AI & Society.

Bursztyn, L., & Jensen, R. (2017). Social image and economic behavior in the field: Identifying, understanding, and shaping social pressure. *Annual Review of Economics*, *9*(1), 131-153.

Carlo, J. L., Lyytinen, K., & Boland Jr, R. J. (2012). Dialectics of collective minding: contradictory appropriations of information technology in a high-risk project. *MIS Quarterly,* 36(4), 1081-1108.

Davis, F. D. (1989). Technology acceptance model: TAM. *Al-Suqri, MN, Al-Aufi, AS: Information Seeking Behavior and Technology Adoption*, *205*, 219.

Davis, F. D., Bagozzi, R. P., & Warshaw, P. R. (1989). User acceptance of computer technology: A comparison of two theoretical models. *Management Science*, 35(8), 982-1003.

Deloitte. (2024). Technology Trust Ethics: Leadership, governance, and workforce decision-making about ethical AI. Deloitte Insights.

Djelic, M. L., & Quack, S. (2007). Overcoming path dependency: path generation in open systems. *Theory and Society*, 36, 161-186.

Djelic, M. L., & Sahlin-Andersson, K. (Eds.). (2006). Transnational governance: Institutional dynamics of regulation. Cambridge University Press.

El Amri, D., & Akrout, H. (2020). Perceived design affordance of new products: Scale development and validation. *Journal of Business Research*, *121*, 127-141.

European Union. (2024). Regulation (EU) 2024/1689 of the European Parliament and of the Council laying down harmonised rules on artificial intelligence (AI Act). *Official Journal of the European Union*, L168, 12 July 2024.

Fazelpour, S., & Danks, D. (2021). Algorithmic bias: Senses, sources, solutions. *Philosophy Compass*, 16(8), e12760. https://doi.org/10.1111/phc3.12760

Floridi, L., & Cowls, J. (2019). A unified framework of five principles for AI in society. *Harvard Data Science Review*, 1(1).

Fountaine, T., McCarthy, B., & Saleh, T. (2019). Building the AI-powered organization. *Harvard Business Review*, *97*(4), 62-73.

Friedrich, J., Kunkel, T., & Thiemann, M. (2024). Becoming influential: Strategies of control, expertise, and socialisation in transnational governance of accounting regulation. *Accounting, Organizations and Society*, 113, 101566.

Gao, T. T., Rohm, A. J., Sultan, F., & Pagani, M. (2013). Consumers un-tethered: A three-market empirical study of consumers' mobile marketing acceptance. *Journal of Business Research*, *66*(12), 2536-2544.

Gegenhuber, T., Logue, D., Hinings, C. B., & Barrett, M. (2022). Institutional perspectives on digital transformation. In *Digital Transformation and Institutional Theory* (Vol. 83, pp. 1-32). Emerald Publishing Limited.

Hafezieh, N., & Eshraghian, F. (2017, June). Affordance theory in social media research: systematic review and synthesis of the literature. In *25th European Conference on Information Systems (ECIS 2017)*.

Hafezieh, N., & Eshraghian, F. (2022). Adopting a 'Search' Lens in Exploration of How Organizations Transform Digitally. In *Proceedings of the 2022 European Conference on Information Systems.* Association of Information Systems.

Hafezieh, N., & Pollock, N. (2023). Digital consumers and the new 'search' practices of born digital organizations. *Information and Organization*, 33(4), 100489.

Jobin, A., Ienca, M., & Vayena, E. (2019). The global landscape of AI ethics guidelines. *Nature Machine Intelligence*, 1, 389-399. https://doi.org/10.1038/s42256-019-0088-2

Johnson, G. (2020). Algorithmic bias: on the implicit biases of social technology. *Synthese*, 198, 9941-9961.

Karimi, J., & Walter, Z. (2015). The role of dynamic capabilities in responding to digital disruption: a factor-based study of the newspaper industry. *Journal of Management Information Systems*, 32(1), 39-81.

Kelly, S., Kaye, S. A., & Oviedo-Trespalacios, O. (2023). What factors contribute to the acceptance of artificial intelligence? A systematic review. *Telematics and Informatics*, *77*, 101925.

Khurana, I., Dutta, D. K., & Ghura, A. S. (2022). SMEs and digital transformation during a crisis: The emergence of resilience as a second-order dynamic capability in an entrepreneurial ecosystem. *Journal of Business Research*, *150*, 623-641.

Kim, C., Mirusmonov, M., & Lee, I. (2010). An empirical examination of factors influencing the intention to use mobile payment. *Computers in Human Behavior*, 26(3), 310-322.

Ko, Y., & Leem, C. S. (2021). The influence of AI technology acceptance and ethical awareness towards intention to use. *Journal of Digital Convergence, 19*, 217-225.

Koenig-Lewis, N., Marquet, M., Palmer, A., & Zhao, A. L. (2015). Enjoyment and social influence: predicting mobile payment adoption. *The Service Industries Journal*, 35(10), 537-554.

Latour, B. (1987). *Science in Action: How to Follow Scientists and Engineers through Society.* Harvard University Press.

Latour, B. (2004). Why has critique run out of steam? From matters of fact to matters of concern. *Critical Inquiry, 30*(2), 225-248

Latour, B. (2005). Reassembling the Social: An Introduction to Actor-Network Theory. Oxford University Press.

Lee, M. J., Pak, A., & Roh, T. (2024). The interplay of institutional pressures, digitalization capability, environmental, social, and governance strategy, and triple bottom line performance: A moderated mediation model. *Business Strategy and the Environment*, 33(6), 5247-5268.

Lobschat, L., Mueller, B., Eggers, F., Brandimarte, L., Diefenbach, S., Kroschke, M., & Wirtz, J. (2021). Corporate digital responsibility. *Journal of Business Research*, *122*, 875-888.

Lu, Y., Yang, S., Chau, P. Y., & Cao, Y. (2011). Dynamics between the trust transfer process and intention to use mobile payment services: A cross-environment perspective. *Information & Management*, 48(8), 393-403.

Mallat, N., Rossi, M., Tuunainen, V. K., & Öörni, A. (2009). The impact of use context on mobile services acceptance: The case of mobile ticketing. *Information & Management*, 46(3), 190-195.

Matt, C., Hess, T., & Benlian, A. (2015). Digital transformation strategies. *Business & Information Systems Engineering*, 57(5), 339-343.

Matthias, A. (2004). The responsibility gap: Ascribing responsibility for the actions of learning automata. *Ethics and Information Technology*, 6(3), 175-183.

Mehrpouya, A., & Salles-Djelic, M. L. (2019). Seeing like the market; exploring the mutual rise of transparency and accounting in transnational economic and market governance. *Accounting, Organizations and Society*, 76, 12-31.

Nambisan, S., Lyytinen, K., Majchrzak, A., & Song, M. (2017). Digital innovation management: Reinventing innovation management research in a digital world. *MIS Quarterly*, 41(1), 223-238.

Nissenbaum, H. (2010). Privacy in Context: Technology, Policy, and the Integrity of Social Life. Stanford University Press.

OECD. (2021). State of implementation of the OECD AI Principles: Insights from national AI policies. *OECD Digital Economy Papers*, No. 311. OECD Publishing.

Olan, F., Arakpogun, E. O., Suklan, J., Nakpodia, F., Damij, N., & Jayawickrama, U. (2022). Artificial intelligence and knowledge sharing: Contributing factors to organizational performance. *Journal of Business Research*, *145*, 605-615.

Pan, Y. C., Jacobs, A., Tan, C., & Askool, S. (2018). Extending technology acceptance model for proximity mobile payment via organizational semiotics. In K. Liu, K. Nakata, W. Li, & C. Baranauskas (Eds.), *Digitalisation, Innovation, and Transformation: 18th IFIP WG 8.1 International Conference on Informatics and Semiotics in Organisations, ICISO 2018* (pp. 43-52). Springer International Publishing.

Power, M. (2015). How accounting begins: Object formation and the accretion of infrastructure. *Accounting, organizations and society*, *47*, 43-55.

Rivard, S. (2004). Information technology and organizational transformation: Solving the management puzzle. Routledge.

Roger, C., & Dauvergne, P. (2016). The rise of transnational governance as a field of study. *International Studies Review*, *18*(3), 415-437.

Saarikko, T., Westergren, U. H., & Blomquist, T. (2020). Digital transformation: Five recommendations for the digitally conscious firm. *Business Horizons*, 63(6), 825-839.

Sanderson, C., Douglas, D., & Lu, Q. (2023). Implementing responsible AI: Tensions and trade-offs between ethics aspects. arXiv preprint

Selander, L., & Jarvenpaa, S. L. (2016). Digital action repertoires and transforming a social movement organization. *MIS Quarterly*, 40(2), 331-352.

Shin, D. H. (2010). Modeling the interaction of users and mobile payment system: Conceptual framework. *International Journal of Human-Computer Interaction*, 26(10), 917-940.

Silva, P. (2015). Davis' technology acceptance model (TAM)(1989). *Information seeking behavior and technology adoption: Theories and trends*, 205-219.

Slade, E. L., Williams, M. D., & Dwivedi, Y. K. (2013). Mobile payment adoption: Classification and review of the extant literature. *The Marketing Review*, 13(2), 167-190.

Song, J., Lee, K., & Khanna, T. (2016). Dynamic capabilities at Samsung: Optimizing internal co-opetition. *California Management Review*, *58*(4), 118-140.

Svahn, F., Mathiassen, L., & Lindgren, R. (2017). Embracing digital innovation in incumbent firms: How Volvo Cars managed competing concerns. *MIS Quarterly*, 41(1), 239-253.

Tan, C., Abdaless, S., & Liu, K. (2018). Norm-based abduction process (NAP) in developing information architecture. In K. Liu, K. Nakata, W. Li, & C. Baranauskas (Eds.), *Digitalisation, Innovation, and Transformation: 18th IFIP*

*WG 8.1 International Conference on Informatics and Semiotics in Organisations, ICISO 2018* (pp. 33-42). Springer International Publishing.

Tan, C., Liu, K., & White, E. (2013). Information architecture for healthcare organizations: the case of a NHS hospital in UK. In Proceedings of the 34th International Conference on Information Systems.

Toon, N. (2024). *How AI Thinks: How we built it, how it can help us, and how we can control it*. Random House.

UK Government. (2023). A pro-innovation approach to AI regulation: White Paper. Department for Science, Innovation and Technology.

Usman, F. O., Eyo-Udo, N. L., Etukudoh, E. A., Odonkor, B., Ibeh, C. V., & Adegbola, A. (2024). A critical review of ai-driven strategies for entrepreneurial success. *International Journal of Management & Entrepreneurship Research*, *6*(1), 200-215.

Vial, G. (2021). Understanding digital transformation: A review and a research agenda. In S. Nambisan (Ed.), *Managing Digital Transformation* (pp. 13-66). Springer.

World Bank Group. (2024). *Global trends in AI governance: Evolving country approaches*. Washington, DC: World Bank.

Zednik, C. (2021). Solving the Black Box Problem: A Normative Framework for Explainable Artificial Intelligence. *Philosophy & Technology*, 34, 265-288.

Zhu, K., Dong, S., Xu, S. X., & Kraemer, K. L. (2006). Innovation diffusion in global contexts: determinants of post-adoption digital transformation of European companies. *European journal of information systems*, *15*(6), 601-616.